\documentstyle[aps,pre,manuscript]{revtex}   

\begin{document}

\draft
\title{Fluctuation-dissipation relationship in chaotic dynamics}

\author{Bidhan Chandra Bag and 
Deb Shankar Ray {\footnote {e-mail : pcdsr@mahendra.iacs.res.in} }
}

\address{Indian Association for the Cultivation of Science,
Jadavpur, Calcutta 700 032, INDIA.}

\maketitle

\begin{abstract}
We consider a general N-degree-of-freedom dissipative 
system which admits of chaotic behaviour. 
Based on a Fokker-Planck description associated with the dynamics
we establish that the drift and the diffusion coefficients
can be related through a set of stochastic parameters which
characterize the steady state of the dynamical system in a way similar to
fluctuation-dissipation relation in non-equilibrium statistical mechanics.
The proposed relationship is verified by numerical experiments on a
driven double well system.
\end{abstract}

\vspace{2.5cm}


\newpage

\section{Introduction}

Although deterministic in principle classically chaotic motion is stochastic
in nature. Ever since the early numerical study of Chirikov mapping \cite{cas}
revealed that the motion of a phase space variable can be characterized by
a simple random walk diffusion equation, attempts have been made to describe 
the chaotic motion in terms of Langevin or Fokker-Planck equations \cite{cas,lic}. 
It is therefore
easy to comprehend a close connection between classical chaos and statistical
mechanics. Two distinct situations arise in this context. The first one
concerns whether classical chaos may serve as a basis for classical 
statistical  mechanics since the ultimate justification of the postulates
of statistical mechanics like Boltzmann hypothesis of molecular chaos, ergodicity or
the postulate of equal a priori probability rests on the dynamics of 
each particle \cite{jkb,ma,gas}. The second one concerns the following : Given that the classical 
chaotic motion is stochastic how and to what extent one can realize the
formulation of statistical mechanics  
for useful description of classical chaos 
\cite{kai,ono,wid,ber,gras,bon,bon1,zur,pat,cohen,pat1,sc1,sc2,sc3,bb1,bb2}
keeping in mind that one essentially deals here with a few-degree-of-freedom 
system. The present paper addresses the second issue.

The emergence of stochastic behaviour of the classically chaotic system is 
due to the loss of correlation of initially nearby trajectories. This is
reflected in the nature of the largest Lyapunov exponent \cite{ben} whose calculation rests
on the linear equation of motion for the separation of these trajectories.
When chaos has fully set in, the time dependence of the linear stability matrix
or Jacobian of the system \cite{fox} in the equation of motion in the tangent space can be
described as a stochastic process since the phase space variables behave as 
stochastic variables. In a number of recent studies we have shown 
\cite{sc1,sc2,sc3,bb1,bb2} that this 
fluctuation of the Jacobian  is amenable to a theoretical description in
terms of the theory of multiplicative noise. This allows us to realize a number 
of important results of nonequilibrium statistical mechanics, like
Kubo relation \cite{sc1}, fluctuation-decoherence relation \cite{sc2}, exponential divergence
of quantum fluctuations \cite{sc3,bb1,bb2}, thermodynamically inspired quantities, e. g. ,
entropy production in chaotic dynamics. Based on a Fokker-Planck description in the tangent space
where the drift and the diffusion coefficients explicitly depend on the phase space
variables or dynamical properties of the system, we show that a connection
between the two moments in terms of the stochastic parameters which 
characterize the long time limit of the dynamical system can be established
in the spirit of fluctuation-dissipation relation. 
We verify the theoretical
proposition by numerical experiments on a simple dissipative 
system.

The rest of the paper is organized as follows : In Sec. II we introduce a
Fokker-Planck description of the dynamical system in the tangent space
and identify the drift and diffusion coefficients which are the functions
of fluctuations of the phase space variables. This is followed by solving the 
Fokker-Planck equation for the steady state distribution required for the calculation
of long time averages in Sec III. In Sec. IV the dynamical stochastic parameters
which characterize the long time behaviour of the system are introduced. The first one of them is
a well-known stochastic parameter closely related to Kolmogorov entropy. 
With the help of these stochastic parameters we 
establish a connection between the drift and diffusion coefficients of the Fokker-Planck
equation in the spirit of
fluctuation-dissipation relation in nonequilibrium statistical mechanics.
In Sec. V we illustrate the general method by an explicit numerical
example to verify the
theoretical proposition. The paper is concluded in Sec. VI.

\section{A Fokker-Planck equation for dissipative 
chaotic dynamics}

We are concerned here with 
a general N-degree-of-freedom system whose Hamiltonian is
given by
\begin{equation}
H = \sum_{i=1}^N \frac{p_i^2}{2 m_i} +V(\{q_i\}, t) \; \;,
\; \; i = 1 \cdots N  \label{e1}
\end{equation}
where $\{q_i, p_i\}$ are the co-ordinate and 
momentum of the i-th degree-of-freedom, respectively, 
which satisfy the generic form of equations 

\begin{equation}
\dot {q}_i = \frac{\partial H}{\partial p_i}  \; \; {\rm and} \; \; 
\dot {p}_i = -\frac{\partial H}{\partial q_i}  \; \;.
\label{e2}
\end{equation}

We now make the Hamiltonian system dissipative by introducing $-\gamma p_i$
on the right hand side of 
the second of Eqs.(2). 
For simplicity we assume $\gamma$ to be the
same for all the N degrees of freedom. 
By invoking the 
symplectic structure of the 
Hamiltonian dynamics as
\begin{eqnarray*}
z_i =
\left\{ \begin{array}{ll}
q_i & \; \; {\rm for} \; i=1 \cdots N  \; \;, \\
p_{i-N} & \; \; {\rm for} \; i=N+1, \cdots 2N \; \;.
\end{array}
\right.
\end{eqnarray*}
and defining I as 
\begin{eqnarray*}
I =
\left[
\begin{array}{cc}
0 & E \\
-E & -\gamma E
\end{array}
\right]
\end{eqnarray*}
where E is an $N\times N$ unit matrix, and $0$ is an  $N\times N$ null matrix,
the equation of motion for the dissipative system can be written 
as
\begin{equation}
\dot{z}_i = \sum_{j=1}^{2N} 
I_{ij} \frac{\partial H}{\partial z_j} \hspace{0.1cm}. \label{e3}
\end{equation}

We now consider two nearby trajectories, $z_i, \dot{z}_i$ and
$z_i+ X_i$, $\dot{z}_i+ \dot{X}_i$ at the same time $t$
in $2N$ dimensional phase space. The time evolution of separation of these 
trajectories is then determined by
\begin{equation}
\dot{X}_i = \sum_{j=1}^{2N} J_{ij}(t) X_j \label{e4}
\end{equation}
in the tangent space $\{X_i\}$, where
\begin{eqnarray}
J_{ij}= \sum_k I_{ik} \frac{\partial^2 H}{\partial z_k \partial z_j}\; \; .
\label{e5}
\end{eqnarray}
Therefore the $2N \times 2N$ linear stability 
matrix $\underline{J}$ assumes the following form
\begin{equation}
\underline{J} =
\left[
\begin{array}{cc}
0 & E \\
\underline{M(t)} & -\gamma E
\end{array}
\right]
\label{e6}
\end{equation}
where \underline{M} is an $N\times N$ matrix. Note that the time dependence of 
stability matrix $\underline{J}(t)$ is due to the second derivative 
$\frac{\partial^2 H}{\partial z_k \partial z_j}$ which is determined \cite{fox} by 
the equation of motion (3). 
The procedure for calculation of $X_i$
and the related quantities is to solve the 
trajectory equation (3) simultaneously
with Eq.(4). Thus when the dissipative 
system described by 
(3) is chaotic, \underline{J}(t) becomes 
(deterministically) stochastic 
due to the fact that $z_i$-s behave as stochastic variables
and the equation of motion (\ref{e4}) in the tangent space
can be interpreted as a stochastic equation \cite{sc1,sc2,sc3,bb1,bb2}.

In the next step we shall be concerned with a stochastic description of 
$\underline{J}(t)$ or $\underline{M(t)}$. For convenience we split up 
\underline{M} into two parts as
\begin{equation}
\underline{M} = \underline{M_0} + \underline{M_1(t)} \label{e7}
\end{equation}
where $\underline{M_0}$ is independent of variables $\{z_i\}$ 
and therefore behaves a sure or constant part
and $\underline{M_1}$ is determined by the
variables $\{z_i\}$ for $i = 1 \cdots 2N$. 
$\underline{M_1}$ refers to the fluctuating part.
We now rewrite the equation of motion (\ref{e4}) in 
tangent space as
\begin{eqnarray}
\dot{X} & = & \underline{J} X  \nonumber \\
       & = & L\left(\left\{X_i\right\}, \left\{z_i\right\}\right)
       \label{e8}
\end{eqnarray}
where $X$ and $L$ are the vectors with $2N$ components. Corresponding to 
(\ref{e7}) $L$ in (\ref{e8}) can be split up again to yield

\begin{eqnarray}
\dot{X} = L^0(X) + L^1(X, \{z_i(t)\})   
\; \; , \; \; \; \; i = 1 \cdots 2N \; \; .
\label{e9}
\end{eqnarray}

Eq.(\ref{e4}) indicates that Eq.(\ref{e8}) 
is linear in $\{X_i\}$. Eqs.(\ref{e4}), (\ref{e5}) and (\ref{e6})
express the fact the first $N$ components of $L^1$ are zero and the last $N$ 
components of $L^1$ are the functions of $\{X_i\}$ for $i = 1 \cdots N$.
The fluctuation in $L_i^1$ is caused by the chaotic variables 
$\left\{z_i\right\}$-s.
This allows us to write the following relation (which will be used later on),
\begin{equation}
\nabla_X \cdot L^1 \phi(\{X_i\}) = L^1 \cdot \nabla_X \phi(\{X_i\}) \; \;
\label{e10}
\end{equation}
where $\phi(\{X_i\}$) is any function of $\{X_i\}$. $\nabla_X$ refers to
differentiation with respect to components $\{X_i\}$ (explicitly 
$X_i = \Delta q_i$ for $i = 1 \cdots N$ and  $X_i = \Delta p_i$ for 
$i = N+1 \cdots 2N$).

Note that Eq.(\ref{e9}) by virtue of (\ref{e8}) 
is a linear stochastic differential equation
with multiplicative noise where the noise is due to $\{z_i\}$ determined by 
equation of motion (\ref{e3}). 
This is the starting point of our further analysis.

Eq.(\ref{e9}) determines a stochastic process with 
some given initial conditions
$\{X_i(0)\}$. We now consider the motion of a representative point $X$ in
$2N$ dimensional tangent space ($X_1 \cdots X_{2N}$) as governed by 
Eq.(\ref{e9}).
The equation of continuity which expresses the conservation of points
determines the variation of density function $\phi(X, t)$ in time as given by
\begin{equation}
\frac{\partial \phi(X, t)}{\partial t} = - {\bf \nabla_X} \cdot L(t) 
\phi(X, t) \; \;.
\label{e11}
\end{equation}

Expressing $A_0$ and $A_1$ as
\begin{equation}
A_0  =  - \nabla_X \cdot L^0 \; \; {\rm and} \; \; 
A_1 =  - \nabla_X \cdot L^1 
\label{e12}
\end{equation}
we may rewrite the equation of continuity as
\begin{equation}
\frac{\partial \phi(X, t)}{\partial t} = [A_0 + \alpha A_1(t)] 
\phi(X, t) \; \;.
\label{e13}
\end{equation}

It is easy to recognize that while $A_0$ denotes the sure part $A_1$ contains the 
multiplicative fluctuations through $\{z_i(t)\}$. $\alpha$ is a parameter
introduced from outside to keep track of the order of fluctuations in the 
calculations. At the end we put $\alpha =1$. 

One of the main results for the linear equations of the form with 
multiplicative noise may now be in order \cite{van}. The average equation of 
$\langle \phi \rangle$ obeys [ $P(x, t) \equiv \langle \phi \rangle$],
\begin{equation}
{\dot P} = \left \{ A_0 + \alpha \langle A_1 \rangle
+ \alpha^2 \int_0^\infty d\tau
\langle \langle A_1(t) \exp(\tau A_0) A_1(t-\tau) 
\rangle \rangle \exp (-\tau A_0) \right \} P(x, t) \; \; .
\label{e14}
\end{equation}

The above result is based on second order cumulant expansion and is 
valid when fluctuations are small but rapid and the correlation 
time $\tau_c$ is short but finite or more precisely

\begin{equation}
\langle \langle A_1(t) A_1(t') \rangle \rangle = 0 \; \; 
\; \; \; {\rm for} \; \; 
\left|t-t'\right| \; > \; \tau_c
\label{e15}
\end{equation}
We have, in general, $\langle A_1 \rangle \ne 0$. Here  
$\langle \langle \cdots \rangle \rangle$ implies
$\langle \langle \zeta_i \zeta_j \rangle \rangle 
= \langle \zeta_i \zeta_j \rangle -\langle \zeta_i \rangle \langle \zeta_j 
\rangle $.

The Eq.(\ref{e14}) is exact in the limit $\tau_c \rightarrow 0$. 
Making use of relation (\ref{e12}) in (\ref{e14}) we obtain 
\begin{eqnarray}
\frac{\partial P}{\partial t} & = & 
\left\{ -\nabla \cdot  L^0  -\alpha \langle \nabla \cdot  L^1 \rangle
+ \alpha^2 \int_0^\infty d\tau
\langle \langle \nabla \cdot L^1(t)   \exp ( -\tau \nabla \cdot L^0)  
\right.\nonumber\\
& & \left. \nabla \cdot L^1(t-\tau) \rangle \rangle
\exp ( \tau \nabla \cdot L^0 )  \right\} P \; \; .
\label{e16}
\end{eqnarray}
The above equation can be transformed into the following Fokker-Planck
equation ($\alpha = 1$) for probability density function $P(X,t)$, 
(the details are given in the Appendix A);
\begin{equation}
\frac{\partial P(X,t)}{\partial t} = -\nabla . F P +
\sum_{i,j} {\cal D}_{ij} \frac{\partial^2 P}{\partial X_i \partial X_j}
\label{e17}
\end{equation}
where,
\begin{equation}
F=L^0 + \langle L^1 \rangle + Q
\label{e18}
\end{equation}
and $Q$ is a $2N$-dimensional vector whose components are
defined by
\begin{eqnarray}
Q_j=-\int_0^\infty \langle \langle R'_j \rangle \rangle
d \tau d_1(\tau) d_2(\tau)
\label{e19}
\end{eqnarray}
Here the determinants $Det_1(\tau)$, $Det_2(\tau)$ and $R'_j$ are given by
\begin{eqnarray}
Det_1(\tau)& = & \left|\frac{d X^{-\tau}}{d X} \right|
\; \; \; \; {\rm and} \; \; Det_2(\tau)=
\left|\frac{d X}{d X^{-\tau}} \right|
\nonumber \\
{\rm and} \; \; \; R'_j & = & \sum_i L_i^1(X,t) \frac{\partial}{\partial X_i}
\sum_k L_k^1(X^{-\tau},t-\tau) \frac{\partial X_j}{\partial X_k^{-\tau}}
\; \; .
\label{a20}
\end{eqnarray}

It is easy to recognize $F$ as an evolution operator. Because 
of the dissipative perturbation we note that div $F < 0$.

The diffusion coefficient ${\cal D}_{ij}$ in Eq.(\ref{e17}) is defined as
\begin{equation}
{\cal D}_{ij}=\int_0^\infty \sum_k \langle \langle L_i^1(X,t) 
L_k^1(X^{-\tau},t-\tau)\frac{dX_j}{dX_k^{-\tau}}
\rangle \rangle Det_1(\tau) Det_2(\tau) d\tau
\label{e21}
\end{equation}
We have followed closely van Kampen's approach \cite{van} to generalized Fokker-Planck
equation (\ref{e17}). Before concluding this section several critical
remarks regarding this derivation need attention:

First, the stochastic process $\underline{M_1(t)}$ determined by $\{z_i\}$ is
obtained {\it exactly} by solving equations of motion (\ref{e3}) for the
chaotic motion of the system. It is therefore necessary to emphasize that we have 
{\it not assumed} any special property of noise, such as, $\underline{M_1(t)}$
is Gaussian or $\delta$-correlated. We reiterate Van Kampen's
emphasis in this approach.

Second, the only assumption made about the noise is that its
correlation time $\tau_c$ is short but finite compared to the
coarse-grained timescale over which the average quantities evolve.

Third, we take care of fluctuations upto second order which implies 
that the deterministic noise is not too strong.

Eq.(\ref{e17}) is the required Fokker-Planck equation in the tangent space
$\{X_i\}$. Note that the drift and diffusion coefficients are
determined by the phase space $\{z_i\}$ properties of the chaotic system and
directly depend on the correlation functions of the fluctuations of the
second derivatives of the Hamiltonian (\ref{e5}).

\section{The Steady state distribution and the calculation of averages}

In what follows we shall be concerned with the long time limit of the 
dynamical system. Thus the steady state distribution of the tangent space 
co-ordinates
$X_i(i = 1\cdots 2N)$ are specially relevant for the present
purpose. To make all these co-ordinates  dimensionless 
we use the following transformations in Eq.(17)

\begin{eqnarray*}
\tau^\prime & = & \omega' t  \; \;, \nonumber \\
y_i & = & \frac{X_i}{d_0} \; \; \; {\rm for} \; i = 1 \cdots N  \; \;,
\end{eqnarray*}

\begin{equation}
y_i = \frac{X_i}{\omega' d_0} \; \; \; \; \; 
{\rm for} \; i = N+1 \cdots 2N \; \;,
\end{equation}

\noindent
where $\omega'$ is a 
scaling constant having dimension of reciprocal of time 
(a possible choice is the linearized frequency of the dynamical system) and
$\tau'$ becomes a dimensionless variable. 
$d_0$ is a constant (to be specified later) having the dimension of length.
The resulting Fokker-Planck Eq.(17) reduces to
\begin{equation}
\frac{\partial P (y, \tau')}{\partial \tau'} = - \nabla . F' (y) P
+  \sum_{i,j} {\cal D'}_{ij} (y)  \frac{\partial^2 P}
{\partial y_i \partial y_j} \; \; .
\end{equation}

Note that Eq.(23) is independent of $d_0$ since $F(X)$ is linear in 
$\{X_i\}$ and ${\cal D}$($X$) is quadratic in $\{X_i\}$. 
Next we consider the stationary state of the system ($\frac{\partial P}
{\partial \tau'} = 0$) and make use of the following linear transformation
( with $\alpha_{2N} = 1$ ) 
\begin{equation}
U = \sum_{i=1}^{2N} \alpha_i y_i
\end{equation}
\noindent

\noindent
in Eq.(23) to obtain the equation for steady state probability distribution
$P_s(U)$ :

\begin{equation}
\frac{\partial}{\partial U} \lambda U P_s (U) + {\cal D}_s \frac{\partial^2 P_s}
{\partial U^2} = 0 \; \; .
\end{equation}

\noindent
$\alpha_i$-s ($i = 1 \cdots 2N-1$) are the constants to be determined.
\noindent
Here 
\begin{equation}
\lambda U = - \sum_i \alpha_i F'_{i} (y)
\end{equation}

\noindent
and
\begin{equation}
{\cal D}_s = \sum_{i, j} {\cal D'}_{ij} \alpha_i \alpha_j \; \; ,
\end{equation}
and we disregard the time dependence of ${\cal D'}$ under weak noise
approximation, to treat ${\cal D'}$ as a constant in the usual way.

Putting (24) in (26) and comparing the coefficients of $y_i$ on both sides we
obtain $2N$ algebraic equation (for $\alpha_i......\alpha_{2N-1}$ and
$\lambda$). The set $\{ \alpha_i\}$ and $\lambda$ are therefore known.

The exact steady state solution, $P_s$ has the well known Gaussian form
which is given by

\begin{equation}
P_s(\{y_i\}) = N \exp \left (-\frac{\lambda}{2 {\cal D}_s} \sum_{i, j}
\alpha_i \alpha_j y_i y_j \right ) \; \;,
\end{equation}

\noindent
where $N$ is the normalization constant.
Eq.(28) expresses the probability distribution of 
tangent space co-ordinates of 
the dynamical system in the long time limit. The important relevant quantity
which measures the separation of initially nearby trajectories when the 
system has attained the stationary state can be computed by calculating the
average of $\sum_{i=1}^{2N} y^2_i$. Making use of 
the distribution (28) we obtain
\begin{equation}
\left \langle \sum_{i =1}^{2N} y^2_i \right \rangle = \frac{{\cal D}_s}{\lambda}
\sum_{i = 1}^{2N} \frac{1}{\alpha^2_i}
\end{equation}
Note that the average as calculated above is a function of $D_s$, $\lambda$ and
$\alpha_i$-s which are dependent on the phase space properties of the dynamical 
system.

\section{Stochastic parameters, connection between ${\cal D}_s$ and $\lambda$ ;
Fluctuation-dissipation relation}

Eq.(25) is a steady state Fokker-Planck equation in tangent space
with linear drift and constant
diffusion coefficients where the co-ordinates have been 
expressed as dimensionless variables $\{y_i\}$. $\lambda$ and ${\cal D}_s$
are the first and second moments, respectively, of the underlying stochastic
process. Our objective here is seek for a connection between the two moments. In
standard nonequilibrium statistical mechanics this connection is expressed by the 
fluctuation-dissipation relation through temperature, an equilibrium parameter 
characterizing the equilibrium state. Our approach here is to
follow a somewhat similar procedure.
This implies that we search for the stochastic parameters which characterize 
the long time limit of the nonlinear dynamical system. We show that an 
appropriate relation between ${\cal D}_s$ and $\lambda$ can be established 
through these parameters.

An important parameter proposed many years ago by Casartelli et. al.\cite{tell}
(a precursor for the largest Lyapunov exponent used as a measure of 
regularity or chaoticity of a nonlinear dynamical system) is the long time
average of $\ln \frac{d(t)}{d_0}$ where $d_0$ is the separation of the 
two initially nearby  trajectories and $d(t)$ is the corresponding 
separation at some time $t$. To express $d(t)$ (having dimension of length)
we write $d(t) = [\sum_{i=1}^{N} (X_i)^2 + \sum_{i = N+1}^{2N} (\frac{X_i}{\omega'})^2]^{\frac{1}{2}}$.
$d(t)$ is determined by solving numerically Eqs. (3) and (4) 
simultaneously or their appropriately transformed version 
for the initial condition $z_0$ corresponding to Eq.(3). In going from j-th
to j+1-th step of iteration in course of time evolution 
any of the components of $X$ say $X_i$ has
to be initialized as $X_i^{j0} = \frac{X_i^j}{d_j} d_0$.
This initialization implies that at each step iteration starts with
same magnitude of $d_0$ but the direction of $d_0$ for step j+1 is that of
$d(t)$ for j-th step (considered in terms of the ratio $\frac{X_i^j}{d_j}$). For
a pictorial illustration we refer to Fig.1 of Ref. \cite{ben}.
j-th time of iteration implies $t = j T$ ($j = 1,2 \cdots \infty)$ and $T$ is 
the characteristic time which corresponds to the shortest ensemble averaged period
of nonlinear dynamical system.  Thus following Casartelli et. al. \cite{tell}
a stochastic parameter can be defined by the following time average
of $\ln \frac{d_j}{d_0}$ as

\begin{equation}
\sigma_n (t, z_0, d_0) = \frac{1}{n}\sum_j^n \ln \frac{d_j}{d_0}
\end{equation}

It has been shown \cite{tell} that as $n\rightarrow \infty$ $\sigma_n$ has a 
definite value. For the disordered system it is positive and for the regular 
system it is zero. The difference of $\sigma_n$ from the largest Lyapunov 
exponent is also noteworthy. Our object here is to generalize (30)
by defining the other higher  order moments (higher than the first
$\sigma_{n \rightarrow \infty}$).
To express these quantities we define first

\begin{equation}
\sigma' = {\rm ln} \frac{d(t)}{d_0}
\end{equation}

We now make use  of the
transformation (22) to express $d(t)$ as a dimensionless quantity in terms
of $\sigma'$ as follows:

\begin{equation}
{\rm ln} \sum_{i = 1}^{2N} {y^2} = 2 \sigma' \; \; .
\end{equation}

The method of cumulant expansion on the other hand tells us that the average of the sum of 
$y^2_i$ can be written as 

\begin{equation}
\left \langle \sum_{i =1}^{2N} y^2_i \right \rangle = 
\exp\left (\sum_m A_m \right ) \; \; \; \; \; \; \; m = 1, 2, 3, \cdots
\end{equation}
 
where $A_m$-s result from 
cumulants of the stochastic quantity $2 \sigma'$. $A_m$-s are 
calculated dynamically from the following relations
\begin{eqnarray}
A_1 & = & m_1  \; \; , \; \; 
A_2  =  \frac{1}{2!} [m_2 - m^2_1] \; \; , \; \; 
A_3  =  \frac{1}{3!} [m_3 - 3 m_1 m_2 + 2 m_1^3] \nonumber\\
A_4 & = & \frac{1}{4!} [m_4 - 3 m_2^2 - 4 m_1 m_3 + 12 m_1^2 m_2 - 6 m_1^4]  
\; \; {\rm etc.}
\end{eqnarray}

\noindent
where $m_k = \frac{2^k}{n} \sum_{j=1}^n \left ({\rm ln} \frac{d_j}{d_0}
\right )^k$
[$k = 1, 2, 3, 4, ...$]. 
In the spirit of Ref. \cite{tell} we enquire, whether these moments/cumulants
reach their steady state values in the long time limit.
We have numerically examined the dependence of $m_k$-s
on various parameters. The parameters are $n$, the time, $d_0$, the measure
of initial separation, the characteristic time $T$ (j-th time of iteration 
implies $t = j T, j = 1, 2, \cdots \infty$). Our observation is that the limit
$m_k$ or limit $A_m$ as $n\rightarrow \infty$ seems to exist in all cases. 
We have examined \cite{bb3} these limits for a number of test cases, e. g. ,
for Lorentz system, Henon-Heiles system and others.
In Fig.(1) we exhibit a typical representative long time behaviour of the 
cumulants $A_m (m = 1 \; {\rm to} \; 4)$ 
for a driven double well potential system discussed in the next section. 
It is apparent that they attain their long time
limits as $n \rightarrow \infty$. Secondly, the first two cumulants are
much
higher compared to others The first moment is the stochastic parameter 
defined by Casartelli et. al. \cite{tell} as a quantity closely related to Kolmogorov
entropy. We are therefore led to believe that the quantities $A_m$-s 
characterize the long time limit or the steady state of a dynamical system.

The relations (33) and (29) can now be combined to give
\begin{equation}
{\cal D}_s = \frac{\lambda}{\sum_{i = 1}^{2n}}\frac{1}{\alpha_i^2} 
\exp \left (\sum_m A_m \right ) \; \; .
\end{equation}

The above relation is the central result of this paper. This establishes a 
connection between the drift and the diffusion coefficients of the Fokker-Planck 
equation (25) through the stochastic  parameters characterizing long time behaviour
of the nonlinear dynamical system. It must be emphasized that both the 
drift $\lambda$ and the diffusion ${\cal D}_s$ coefficients arise from 
the deterministic stochasticity implied in the dynamical equation
of motion (3). The relation (35) is therefore reminiscent of the familiar
fluctuation-dissipation relation.

A few points regarding the relation (35) are in order. It is  important to 
note that the fluctuation-dissipation relation in conventional nonequilibrium
statistical mechanics is valid for a stochastic system for which the noise 
is internal. The spiritual root of this relation lies at the dynamic balance 
between the input of energy into the system from the fluctuations of the 
surrounding and the output of energy from the system due to its dissipation
into the surrounding. The system-reservoir model \cite{loui,gg} developed over the last few
decades suggests that the coupling between the system and the reservoir is 
responsible for a common origin of drift and diffusion. In the present theory 
this common mechanism is the fluctuations of the phase space variables (or
second derivative of the Hamiltonian) inherent in both the drift $\lambda$
and the diffusion $D_s$ coefficients of the Fokker-Planck equation. We
point out that the relation is still valid for the pure Hamiltonian system 
($\gamma = 0$). For this reason the relation (35) is somewhat formal in
contrast to the standard fluctuation-dissipation relation.

\section{An example and numerical verification}

To illustrate the theory developed above, we now choose a driven
double-well oscillator system with Hamiltonian
\begin{equation}
H=\frac{p_1^2}{2}+aq_1^4-bq_1^2+ \epsilon q_1\cos\Omega t
\end{equation}
where $p_1$ and $q_1$ are the momentum and position variables of the 
system. $a$ and $b$ are the constants characterizing the potential.
$\epsilon$ includes the effect of coupling constant and the
driving strength of the external field with frequency $\Omega$. This model
has been extensively used in recent years for the study of 
chaotic dynamics \cite{sc1,sc2,lin}.

The dissipative equations of motion for the tangent space variables $X_1$ 
and $X_2$ corresponding to $q_1$ and $p_1$ (Eq.8) read as follows:
\begin{equation}
\frac{d}{d t} \left[
\begin{array}{c}
X_1 \\
X_2
\end{array} \right]
= \underline{J} \left[
\begin{array}{c}
X_1 \\
X_2
\end{array} \right] \; \; , \; \;
\left \{
\begin{array}{c}
\Delta q_1 = X_1 \\
\Delta p_1 = X_2
\end{array} \right \} \; \; .
\label{e40}
\end{equation}
where $\underline{J}$ as expressed in our earlier notation 
$z_1 = q_1$ and $z_2 = p_1$ is given by
\begin{eqnarray*}
\left(
\begin{array}{cc}
0 & 1 \\
\zeta (t)+2b \;  & \; -\gamma
\end{array}
\right) \; \; ,
\end{eqnarray*}
where $\zeta(t)=-12 a z_1^2$. 
Eq.(\ref{e40}) is thus rewritten as 
\begin{equation}
\frac{d}{d t} \left(
\begin{array}{c}
X_1 \\
X_2
\end{array}
\right) = L^0+L^1
\label{e41}
\end{equation}
with
\begin{eqnarray*}
L^0=\left(
\begin{array}{c}
X_2\\
2bX_1-\gamma X_2
\end{array}
\right)
\; \; \; \; {\rm and} \; \;
L^1=\left(
\begin{array}{c}
0\\
\zeta (t) X_1
\end{array}
\right) \; \; ,
\end{eqnarray*}
where $L^0$ and $L^1$ are the constant and the 
fluctuating parts(vectors), respectively.
The fluctuation in $L^1$, i.e., in $\zeta(t)$ is due to stochasticity of the
following chaotic dissipative dynamical equations of motion;
\begin{equation}
\dot{z}_1 = z_2 \; \; {\rm and} \; \; 
\dot{z}_2  =  -az_1^3+2bz_1-\epsilon \cos\Omega t -\gamma z_2 \; \; .
\label{42}
\end{equation}
The result of Eq.(\ref{a5}) can then be applied and after some algebra
the Fokker-Planck equation (17) for the dissipative driven double-well
oscillator assumes the following form:
\begin{eqnarray}
\frac{\partial P}{\partial t} & = & -X_2 \frac{\partial P}{\partial X_1}
-\omega^2 X_1 \frac{\partial P}{\partial X_2} + \gamma 
\frac{\partial}{\partial X_2} (X_2 P) +{\cal D}_{21} \frac{\partial^2 P}
{\partial X_2 \partial X_1} + {\cal D}_{22} \frac{\partial^2 P}
{\partial X_2^2}
\end{eqnarray}
\noindent
where 

\begin{eqnarray*}
{\cal D}_{21}=
X_1^2 \int_0^\infty \langle\langle \zeta(t) \zeta(t-\tau) \rangle \rangle
\tau e^{-\gamma \tau}d\tau
\end{eqnarray*}
and
\begin{equation}
{\cal D}_{22}=
X_1^2 \int_0^\infty \langle\langle \zeta(t) \zeta(t-\tau) \rangle \rangle
e^{-\gamma \tau}d\tau 
- X_1 X_2 \int_0^\infty \langle\langle \zeta(t) \zeta(t-\tau) 
\rangle \rangle
\tau e^{-\gamma \tau}d\tau  \; \; 
\end{equation}

\noindent
with
\begin{equation}
\omega^2 =  2b+c+c_2 \; \; , \; \; 
c_2  =  \int_0^\infty \langle\langle \zeta(t)\zeta(t-\tau) \rangle\rangle  
\tau e^{-\gamma \tau} d\tau \; \; {\rm and} \; \;
c  =  \langle \zeta \rangle \; \; .
\end{equation}

\noindent
The similarity of the equation (40) to generalized Kramers' equation can not
be overlooked. This suggests a clear interplay of chaotic diffusive motion
and dissipation in the dynamics.

Using the transformation (22) Eq.(40) can be written as
\begin{eqnarray}
\frac{\partial P}{\partial \tau'} & = & -y_2 \frac{\partial P}{\partial y_1}
-\overline{\omega}^2 y_1 \frac{\partial P}{\partial y_2} + \overline{\gamma} 
\frac{\partial}{\partial y_2} (y_2 P) +
{\cal D'}_{21} \frac{\partial^2 P}
{\partial y_2 \partial y_1} + {\cal D'}_{22} \frac{\partial^2 P}
{\partial y_2^2} 
\end{eqnarray}

\noindent
where 
\begin{eqnarray*}
\overline{\omega}^2 = \frac{\omega^2}{\omega'^2} \; \; , 
\; \; \overline{\gamma} = \frac{\gamma}{\omega'} \; \; , \; \; 
{\cal D'}_{21}= \frac
{y_1^2 (0)}{{\omega'}^2}\int_0^\infty \langle\langle \zeta(\tau') 
\zeta(\tau'-\tau) \rangle \rangle
\tau e^{-\gamma \tau}d\tau \; \; {\rm and}
\end{eqnarray*}

\begin{equation}
{\cal D'}_{22}= \frac
{y_1^2 (0)}{{\omega'}^2} \int_0^\infty \langle\langle \zeta(\tau') 
\zeta(\tau'-\tau) \rangle \rangle
e^{-\gamma \tau}d\tau 
- \frac{y_1(0) y_2 (0)}{\omega'}\int_0^\infty \langle\langle \zeta(\tau') 
\zeta(\tau'-\tau) \rangle \rangle
\tau e^{-\gamma \tau}d\tau  \; \; 
\end{equation}

\noindent
and the time dependence of $y_1$ and $y_2$ in the diffusion coefficients 
have been frozen under weak noise approximation.

Now using the linear transformation (24) in Eq.(43) we obtain in the 
stationary state
\begin{equation}
\frac{\partial}{\partial U} \lambda U P_s +{\cal D}_s \frac{\partial^2 P_s}
{\partial U^2} = 0
\end{equation}

\noindent
where
\begin{equation}
U =  \alpha_1 y_1 + y_2 \; \; {\rm and} \; \;
\lambda U  =  - \alpha_1 y_2 -{\overline{\omega}}^2 y_1 
+ \overline{\gamma} y_2
\end{equation}

\noindent
and
\begin{equation}
{\cal D}_s = {\cal D'}_{22}
\end{equation}

where for simplicity it has been assumed that ${\cal D'}_{21}$
is much small compared to the Markovian contribution ${\cal D'}_{22}$.

Comparing the coefficients of $y_1$ and $y_2$ on both sides of Eq.(46) we obtain
\begin{eqnarray*}
\lambda \alpha_1 = - {\overline{\omega}}^2
\; \; {\rm and} \; \; 
\lambda = - \alpha_1 + \overline{\gamma}
\end{eqnarray*}

\noindent
Therefore we have
\begin{equation}
\alpha_1 = \frac{-\overline{\gamma} - \sqrt{{\overline{\gamma}}^2 + 4 
{\overline{\omega}}^2}}{2} \; \; {\rm and} \; \;
\lambda = \frac{\overline{\gamma} + \sqrt{{\overline{\gamma}}^2 + 4 
{\overline{\omega}}^2}}{2} \; \; .
\end{equation}

Here the negative value of $\alpha_1$ is taken to make $\lambda$ positive
for a physically allowed solution of the steady state distribution (49).
The solution of Eq.(45) is given by
\begin{equation}
P_s  =  N \exp\left(- \frac{\lambda}{2 {\cal D}_s} ({\alpha_1}^2 y_1^2 + 
2 \alpha_1 y_1 y_2 + y_2^2)\right) \; \; .
\end{equation}

With the help of above distribution the average quantities in
tangent space can be calculated. Thus we have
\begin{equation}
\left \langle y_1^2 + y_2^2 \right \rangle = 
\frac{{\cal D}_s}{\lambda} \left ( \frac{1}
{\alpha_1^2} + 1 \right ) \; \; .
\end{equation} 

The fluctuation dissipation relation (35) can then be obtained by combining (50) with 
(33) as follows ;
\begin{equation}
{\cal D}_s = \frac{\lambda}{(\frac{1}{\alpha_1^2} +1 )} 
\exp \left (\sum_m A_m \right) \; \; .
\end{equation}

$\lambda$ and $\alpha_1$ are to be calculated using (48). For these we require 
explicit numerical evaluation of $\overline{\omega}^2$ as defined in (43) and 
(44). 
The dissipative chaotic motion is governed by equations (37) and (39). We
choose the following values of the parameters \cite{lin} $a = 0.5$, $b = 10$, 
$\epsilon = 10$,  
$\Omega = 6.07$ and $\gamma=0.001$. The coupling-cum-field strength $\epsilon$
has been varied from set to set. We choose the initial conditions $z_1(0) = -3.5$ and
$z_2(0) = 0$ which ensures strong global chaos. Note that $c_2$ as expressed
in (42) and in the diffusion coefficients are the integrals over the 
correlations of $\zeta(t)$ ($\zeta(t)$ is the fluctuating part of the second 
derivative of the potential $V(z)$ and is given by $\zeta(t) = -12 az_1^2$). 
To calculate the correlation function $\langle \langle\zeta(t)\zeta(t-\tau)
\rangle\rangle$ and the average $\langle\zeta(t)\rangle$ it is necessary 
to determine long time series in $\zeta(t)$ by numerically solving the classical
equation of motion (39). The next step is to carry out the averaging over
the time series.
For further details of the numerical procedure we refer to the earlier
work \cite{sc3,bb1,bb2}. On the other hand the cumulants $ A_m (m = 1, 2, 3, 4)$ (as defined in (34)
and (35)) are calculated from Eqs.(37) and (39) directly. The method has 
already been outlined in Sec.(IV) and in Ref. \cite{tell}.  We then plot the  theoretically
calculated values of ${\cal D}_s$ from the evaluation of $\lambda$, $\alpha_1$
and the cumulants for several values of the coupling constant $\epsilon$
(Eq.36) and compare them with the diffusion coefficients obtained from 
the direct numerical integration of Eqs.(39) and (37) with the appropriate 
transformation (22) for the corresponding values of $\epsilon$.
The result is shown in Fig. 2. It may be noted that the theoretical and 
numerical results are in good agreement. The validity of the 
fluctuation-dissipation relation as proposed in Eq.(35) is therefore
reasonably satisfactory.

\section{Conclusions}

The crucial question of instability of classical motion essentially rests on
the linear stability matrix or Jacobian matrix associated with the 
equations of motion. While the linear stability analysis
around the fixed points is based on the assumption of constancy of this
matrix we take full account of the time dependence of the quantity in the chaotic regime by 
considering it to be a stochastic process, since the  phase variables 
behave stochastically. Based on a Fokker-Planck description in the tangent space
we trace the origin of chaotic diffusion and drift in the correlation of 
fluctuations of the linear stability matrix.

The main conclusions of this study are the following :

(i) We show that a class of dynamical stochastic parameters which attain their 
steady state values in the long time limit of the dynamical system may be
used to characterize the dynamical steady state of the system. The first one
of them which was proposed by Casartelli et. al. \cite{tell} several years ago as a
measure of the chaoticity of the system is closely related to Kolmogorov
entropy.

(ii) We establish a connection between the drift and the diffusion coefficients of
the Fokker-Planck equation and the dynamical stochastic parameters in the spirit
of fluctuation-dissipation relation. The realization of this relation in chaotic 
dynamics therefore carries the message that although
comprising a few degrees of freedom a chaotic system may behave as a statistical
mechanical system (although in a somewhat different sense).

The theoretical relations proposed here are generic for N-degree-of-freedom 
chaotic Hamiltonian system with or without dissipation and have been verified
by numerical analysis of a driven nonlinear dissipative system. We hope that the present
approach will find useful application in searching for the related thermodynamically
inspired quantities in few-degree-of-freedom systems.

\acknowledgments
B. C. Bag is indebted to the Council of Scientific and
Industrial Research (C.S.I.R.), Government of India, for a fellowship. 

\newpage

\begin{appendix}

\section{The derivation of the Fokker-Planck equation}

We first note that the
operator $e (-\tau \nabla \cdot  L^0)$ provides the solution 
of the equation [Eq.(\ref{e13}), $\alpha = 0$]
\begin{equation}
\frac{\partial f(X, t)}{\partial t} = -\nabla_X \cdot  L^0 f(X,t)
\label{a1}
\end{equation}
$f$ signifies the ``unperturbed'' part of $P$ which can be found 
explicitly in terms of characteristic curves. The equation
\begin{equation}
\dot X = L^0 (X) 
\label{a2}
\end{equation}
determines for a fixed $t$  a mapping from $X(\tau=0)$ to $X(\tau)$, i. e., 
$X \rightarrow X^\tau$ with inverse $(X^\tau)^{-\tau}=X$ .
The solution of (\ref{a1}) is
\begin{equation}
f(X,t)= f(X^{-t}, 0) \left | \frac{d X^{-t}}{d X} \right | = 
e \left [ -t \nabla \cdot F_0  \right ]  f(X, 0 ).
\label{a3}
\end{equation}
$\left | \frac{d(X^{-t})}{d(X)} \right |$ being a Jacobian determinant. The
effect of $e(-t \nabla \cdot  L^0)$ on $f(X)$ is as 
\begin{equation}
e(-t \nabla \cdot L^0) f(X,0) = f(X^{-t}, 0) \left|\frac
{d X^{-t}}{d X} \right| \; \;.
\label{a4}
\end{equation}

This simplification in Eq.(\ref{e16}) yields
\begin{eqnarray}
\frac{\partial P}{\partial t} & = &
\left\{ -\nabla \cdot  L^0 
-\alpha \langle \nabla \cdot L^1 \rangle + \alpha^2 \int_0^\infty
d\tau \left| \frac{d X^{-\tau}}{dX}
\right| \right. \nonumber \\
& & \langle \langle \nabla \cdot  L^1(X,t) 
{\bf \nabla}_{-\tau} \cdot L^1({\bf x}^{-\tau}, t-\tau) \rangle 
\rangle \left. \left| \frac{dX}{d X^{-\tau}}
\right|  \right\} P \; \;.
\label{a5}
\end{eqnarray}

Now to express the Jacobian, ${X}^{-\tau}$ and $\nabla_{-\tau}$
in terms of $\nabla$ and $X$ we solve Eq.(\ref{a2}) 
for short time (this is
consistent with the assumption that the fluctuations are rapid \cite{van}). 

We now write the solution of Eq.(\ref{a2}) [using Eqs.(4-6)] as follows ;

\begin{equation}
\left(
\begin{array}{c}
X_1^{-\tau} \\
\vdots  \\
X_N^{-\tau}
\end{array}
\right)
= -\tau \left(
\begin{array}{c}
X_{N+1}\\
\vdots  \\
X_{2N}
\end{array}
\right)+\left(
\begin{array}{c}
X_1\\
\vdots  \\
X_N
\end{array}
\right)
=\left(
\begin{array}{c}
\bar{G}_1(X)\\
\vdots  \\
\bar{G}_N(X)
\end{array}
\right)
\label{a6}
\end{equation}
and
\begin{equation}
\left(
\begin{array}{c}
X_{N+1}^{-\tau} \\
\vdots  \\
X_{2N}^{-\tau}
\end{array}
\right)
= e^{\gamma \tau} \left(
\begin{array}{c}
X_{N+1}\\
\vdots  \\
X_{2N}
\end{array}
\right)-\tau\left(
\begin{array}{c}
G_{N+1}(X)\\
\vdots  \\
G_{2N}(X)
\end{array}
\right)
=\left(
\begin{array}{c}
\bar{G}_{N+1}(X)\\
\vdots  \\
\bar{G}_{2N}(X)
\end{array}
\right)
\label{a7}
\end{equation}
Here the terms of $O(\tau^2)$ are neglected. Since the 
vector $X^{-\tau}$ is expressible as a function of $X$ we write
\begin{equation}
X^{-\tau} = \bar{G}(X) \; \; ,
\label{a8}
\end{equation}
and the following simplification holds good;
\begin{eqnarray}
L^1(X^{-\tau}, t-\tau).\nabla_{-\tau} & = &
L^1(\bar{G}(X),t-\tau) . \nabla_{-\tau} \nonumber \\
& = & \sum_k L_k^1(\bar{G}(X),t-\tau)\frac{\partial}{\partial X_k^{-\tau}}
\nonumber \\
& = & \sum_j \sum_k L_k^1(\bar{G}(X),t-\tau)g_{jk}
\frac{\partial}{\partial X_j} \; \; \; \; \; 
\; \; \; ; j,k = 1 \cdots 2N
\label{a9}
\end{eqnarray}
where 
\begin{equation}
g_{jk}=\frac{\partial X_j}{\partial X_k^{-\tau}}
\label{a10}
\end{equation}

In view of Eqs.(\ref{a6}) and (\ref{a7}) we note:
\begin{eqnarray*}
{\rm if} \; \; \; j=k \; \; {\rm then} \; \; g_{jk} & = & 1, \; \; k=1 \cdots N \\
& = & e^{-\gamma \tau}, \; \; k=N+1 \cdots 2N\\
\\
{\rm if} \; \; \; j \ne k \; \; {\rm then} 
\; \; g_{jk} & \propto & -\tau e^{-\gamma \tau} \; \; {\rm or} \; \;  0
\end{eqnarray*}

\noindent
Thus $g_{jk}$ is a function of $\tau$ only. Let 
\begin{equation}
R_j = \sum_k L_k^1(\bar{G}(X),t-\tau)g_{jk}
\label{a11}
\end{equation}
From Eqs.(\ref{e8}), (\ref{e9}) and (\ref{a8}) we write
\begin{equation}
L_i^1(X^{-\tau},t-\tau)
=L_i^1(\bar{G}(X),t-\tau) = 0 \; \; \; \; {\rm for} \; 
i=1 \cdots N
\label{a12}
\end{equation}
So the conditions (\ref{a11}), (\ref{a12}) and (\ref{a6}) imply that
\begin{eqnarray}
R_j(X, t-\tau) & = & R_j(X_1 \cdots X_N, t-\tau) 
\; \; \; {\rm for} \; j=1 \cdots N \nonumber \\
R_j(X, t-\tau) & = & R_j(X_1 \cdots X_{2N}, t-\tau) 
\; \; \; {\rm for} \; j=N+1 \cdots 2N
\label{a13}
\end{eqnarray}
We next carry out the following simplifications of 
$\alpha^2$-term in Eq.(\ref{a5}).
We make use of the relation (\ref{e10}) to obtain
\begin{eqnarray}
L^1(X,t).\nabla\sum_jR_j\frac{\partial}{\partial X_j} P(X,t)
& = & \sum_i L_i^1(X,t)\frac{\partial}{\partial X_i}\sum_j R_j
\frac{\partial}{\partial X_j}P(X,t)
\nonumber \\
& = & \sum_{i,j} L_i^1(X,t)
R_j \frac{\partial^2}{\partial X_i \partial X_j} P(X,t)\nonumber \\
&&+\sum_j
R'_j
\frac{\partial}{\partial X_j}P(X,t)
\label{a14}
\end{eqnarray}
where
\begin{equation}
R'_j=\sum_i L_i^1 (X,t) \frac{\partial}{\partial X_i} R_j
\label{a15}
\end{equation}
The conditions (\ref{a12}) and (\ref{a13}) imply that
\begin{eqnarray}
R'_j & = & 0 \; \; \; \; \; {\rm for} \; j=1 \cdots N \nonumber \\
R'_j & = & 
R'_j(X_1 \cdots X_N, t-\tau)  \ne 0 ;\; \; \; \; \; {\rm for} 
\; j=N+1 \cdots 2N 
\label{a16}
\end{eqnarray}
By (\ref{a16}) one has
\begin{equation}
R' . \nabla P(X,t) = \nabla . R' P(X,t)
\label{a17}
\end{equation}

Making use of Eqs.(\ref{e10}), (\ref{a9}), 
(\ref{a14}) and (\ref{a17}) in Eq.(\ref{a5}) we obtain the
Fokker-Planck equation (\ref{e17}).

\end{appendix}

\newpage

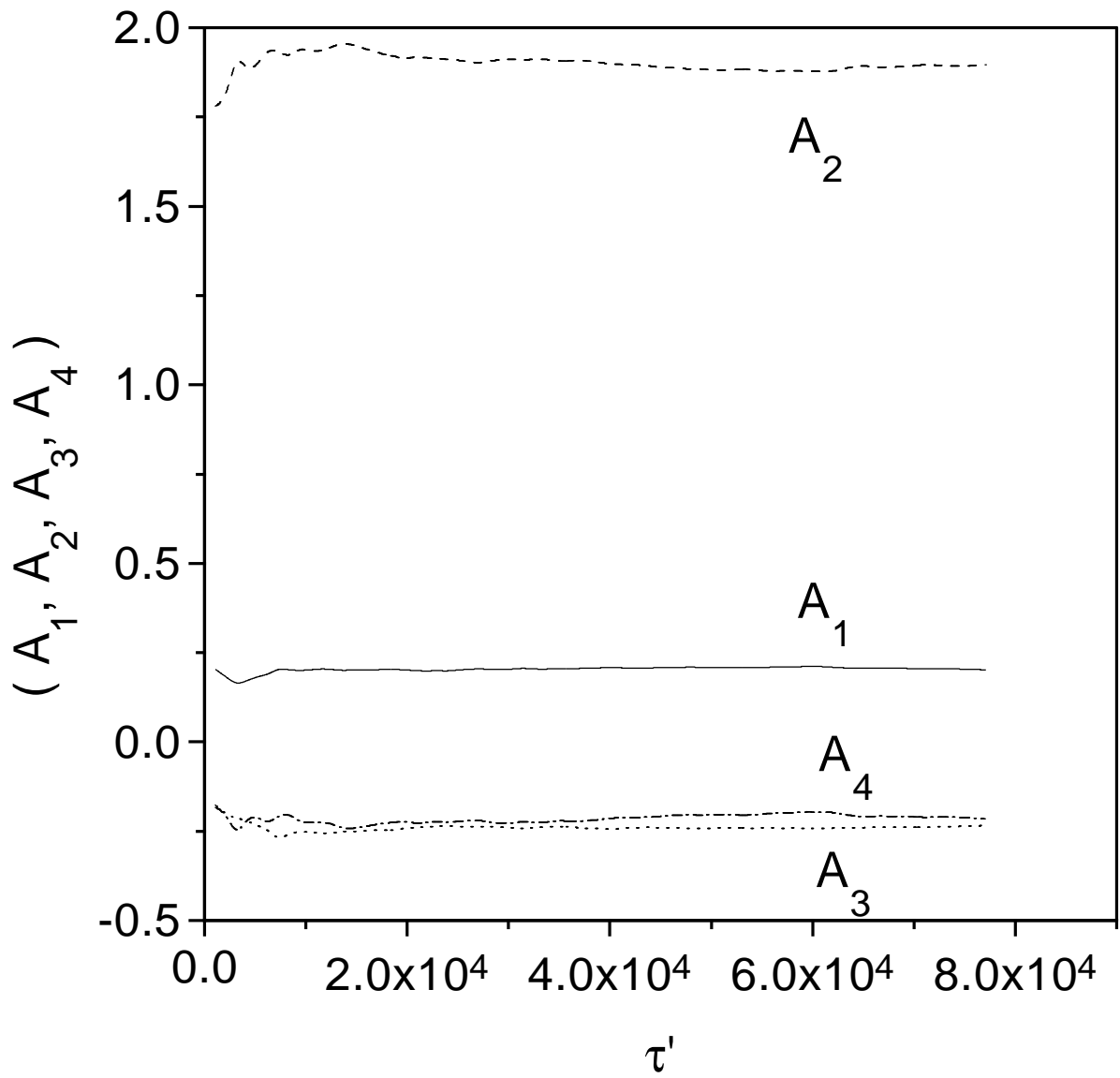
\begin{figure}
\caption{
The first four dimentionless cumulants $A_1, A_2, A_3$ and $A_4$ are plotted 
against dimentionless time for the
dynamical system described by Eq.(39).
}
\end{figure}

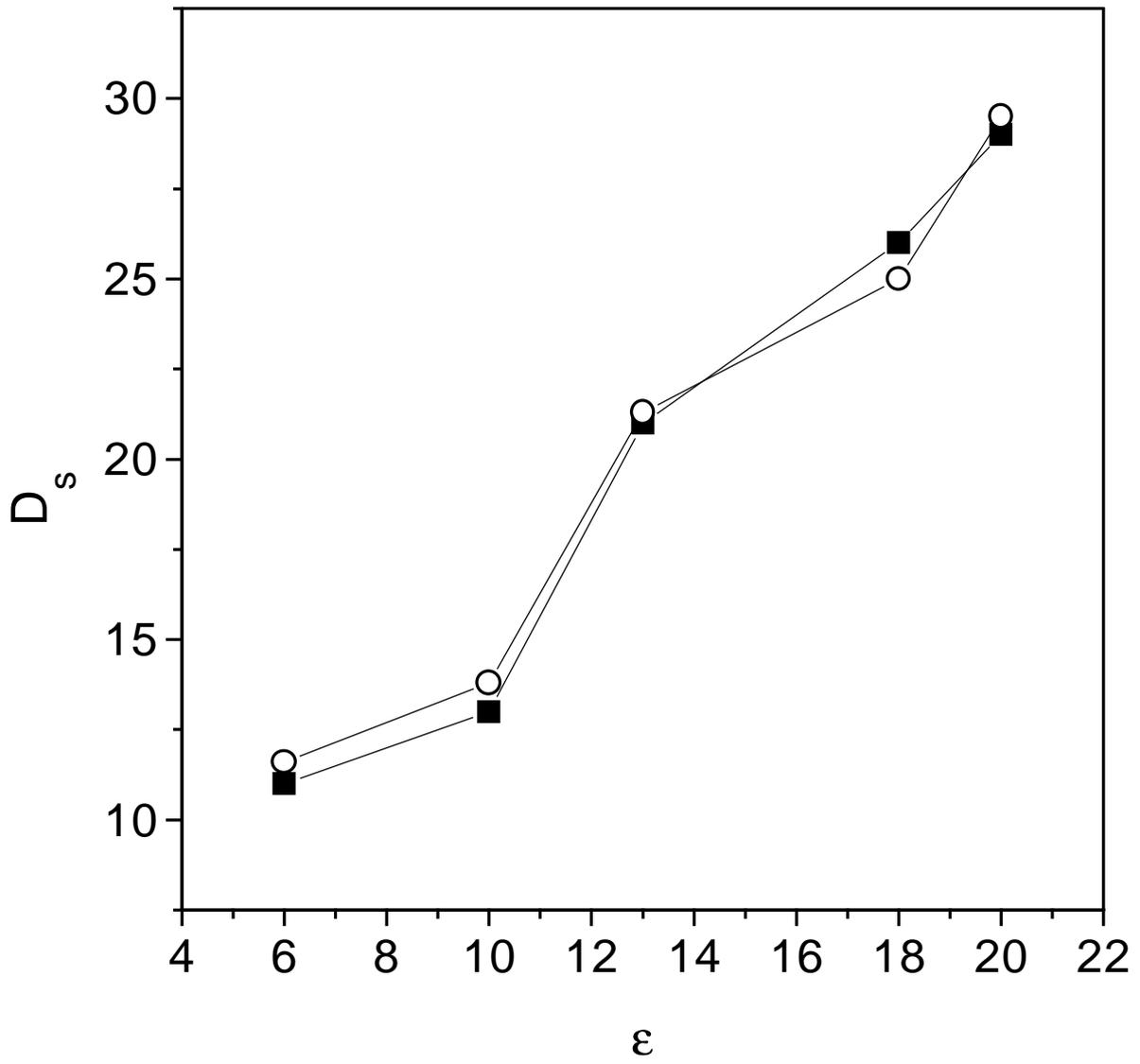
\begin{figure}
\caption{
The diffusion coefficients calculated numerically (marked as dark 
squares) using Eqs. (38) and (39) after transformation (22) are compared
with theoretically obtained values (marked as circles) using Eq.(51) for several values of
the coupling-cum-external field strength $\epsilon$ (units are arbitrary).
}
\end{figure}

\end{document}